\documentclass[sn-nature,iicol]{sn-jnl}
\usepackage{graphicx}%
\usepackage{multirow}%
\usepackage{amsmath,amssymb,amsfonts}%
\usepackage{amsthm}%
\usepackage{mathrsfs}%
\usepackage[title]{appendix}%
\usepackage{xcolor}%
\usepackage{textcomp}%
\usepackage{manyfoot}%
\usepackage{booktabs}%
\usepackage{algorithm}%
\usepackage{algorithmicx}%
\usepackage{algpseudocode}%
\usepackage{listings}%
\usepackage{pifont,xcolor} 
\usepackage{pdfpages}
\usepackage{pgffor}
\theoremstyle{thmstyleone}%
\theoremstyle{thmstyletwo}%
\theoremstyle{thmstylethree}%
\usepackage[dvipsnames]{xcolor}
\usepackage{subfiles}
\usepackage{tikz}
\usepackage{pgfplots}
\usepgfplotslibrary{colorbrewer}
\pgfplotsset{compat=1.18}
\usetikzlibrary{positioning, calc, backgrounds, shapes.geometric, shapes.arrows, arrows.meta, shapes.symbols, decorations.text, fit}
\usetikzlibrary{external}
\usepgfplotslibrary{colormaps}
\usepgfplotslibrary{statistics}
\usepackage[most]{tcolorbox}
\usepackage{hyperref}
\begin{document}
\title[Generative AI for sustainable plant-based foods]
{{\sffamily{\bfseries{Generative AI enables human-centered design 
of sustainable plant-based foods}}}}
\author[1]{\fnm{Vahidullah} \sur{Tac}}\email{vtac@stanford.edu}
\author*[1]{\fnm{Ellen} \sur{Kuhl}}\email{ekuhl@stanford.edu}
\affil[1]{\orgdiv{Department of Mechanical Engineering}, \orgname{Stanford University}, \orgaddress{\city{Stanford}, \country{USA}}}
\abstract{Dietary transitions away from animal-based meat could reduce food-related environmental burdens, but plant-based alternatives must compete on flavor, texture, and overall liking. Here we extend a generative AI framework with 469 plant-based burger recipes and use it to create six formulations across five primary ingredients: bean, chickpea, lentil, mushroom, and oat. An executive chef translates the AI-generated formulations into standardized preparation protocols and refines the ingredient quantities based on culinary expertise. The integration of generative modeling and culinary expertise defines a human-centered generative AI framework for food design. We evaluate the six burgers in a blinded sensory survey with 101 participants in an operating restaurant and quantify their land use, greenhouse gas emissions, eutrophication potential, and scarcity-weighted water use. We use Big Mac scores from a previous survey as an iconic commercial benchmark. The mushroom burger achieves mean flavor, texture, and overall liking scores similar to the Big Mac, with less than one-third of its estimated environmental impact. The other formulations reduce the estimated impact further, in several cases to near one-tenth of the Big Mac value, but receive lower liking scores. Mixed-effects models quantify associations between sensory attributes and human liking and retain predictive accuracy on independent survey data, with mean absolute errors no greater than 0.52 points on a seven-point scale. Together, these results demonstrate how generative design, culinary expertise, human sensory response, and environmental assessment can converge to create sustainable foods that people want to eat.}
\keywords{artificial intelligence $|$ generative AI $|$ machine learning $|$ plant-based foods $|$ sustainable foods}
\maketitle
Food choices shape both human and planetary health \cite{keesing2022}. The global food system contributes substantially to greenhouse gas emissions, land conversion, freshwater use, nutrient pollution, and biodiversity loss \cite{tilmanGlobalDietsLink2014}. Animal agriculture accounts for a disproportionate share of these burdens, and beef generally imposes the largest environmental costs among common animal products \cite{eshelLandIrrigationWater2014, pooreReducingFoodsEnvironmental2018}. High consumption of red and processed meat also correlates with greater risks of cardiovascular disease, type 2 diabetes, and several cancers \cite{shiRedMeatConsumption2023}. Reducing meat consumption could therefore improve public health and environmental sustainability at the same time.

Plant-rich diets offer a promising alternative. Global models show that dietary shifts away from animal products can lower food-related emissions and premature mortality \cite{springmannAnalysisValuationHealth2016}. Comparisons across food groups reveal the same pattern: foods that promote health often require fewer environmental resources \cite{clarkMultipleHealthEnvironmental2019}. In a controlled dietary trial, replacing animal-based meat with plant-based alternatives also improved several cardiovascular risk factors \cite{crimarcoRandomizedCrossoverTrial2020}. These findings identify plant-based foods as a practical substitute for high-impact products such as beef.

Consumer acceptance, however, limits this transition. People consistently rank flavor and texture among the strongest determinants of food choice, and plant-based meat alternatives vary widely in both sensory quality and overall liking \cite{szenderakConsumerAcceptancePlantBased2022, giezenaarProfilingNovelPlantbased2024}. Many products lack the juiciness, meatiness, and savory flavor that consumers expect from meat \cite{st.pierreMeatballMatchupPlant2026}. Environmental or health benefits may encourage consumers to try an alternative, but sensory quality  strongly influences whether they choose it again. Plant-based foods must therefore compete with meat on taste and texture, not only on nutrition or sustainability \cite{datta2026}.

Meeting this requirement poses a complex design problem. Ingredient identities, quantities, and interactions jointly determine a burger’s flavor, texture, nutrition, and environmental impact \cite{tacGenerativeArtificialIntelligence2026}. Conventional product development relies on culinary expertise, iterative reformulation, and repeated sensory tests, which restrict exploration of the vast formulation space \cite{kuhlAIFoodAccelerating2025}. Machine-learning methods can predict sensory outcomes and optimize predefined recipes \cite{zhangFoodProductDesign2019}, while generative models can also propose new ingredient combinations \cite{tacGenerativeAIMaterial2026a}. Together, these capabilities create an opportunity to learn consumer preferences and use them to guide food design.

Here we investigate how generative artificial intelligence can support the human-centered design of plant-based burgers. Building on our previously validated framework \cite{tacGenerativeArtificialIntelligence2026}, we enrich the training data with 469 plant-based recipes and generate six formulations across five primary ingredient systems. We combine generative design with restaurant-based sensory evaluation, environmental assessment, and mixed-effects modeling to learn how perceived flavor and texture relate to human liking. Among the generated formulations, the mushroom burger achieves mean flavor, texture, and overall liking scores similar to the Big Mac while producing less than one-third of its environmental impact, and the learned sensory relationships remain predictive in independent survey data. Together, these results establish a quantitative framework that links generative food design to human sensory preference and environmental sustainability.
\section*{\sffamily{\bfseries{Results}}}
We extended our generative AI framework to design plant-based burgers with bean, chickpea, lentil, mushroom, or oat as the primary ingredient. We selected these primary ingredients from the most prevalent primary ingredients in the training database of 2,679 burger recipes. For the oat burger, we selected two formulations, one savory and one sweet. Overall, we generated six formulations: bean, chickpea, lentil, mushroom, savory oat, and sweet oat burgers. An executive chef translated these formulations into standardized preparation protocols and refined the ingredient quantities while preserving the AI-generated ingredient selections. The Supplementary Information provides the complete recipes for all six burgers.

\begin{figure*}[t]
\centering
\includegraphics[width=13cm,height=7.5cm]{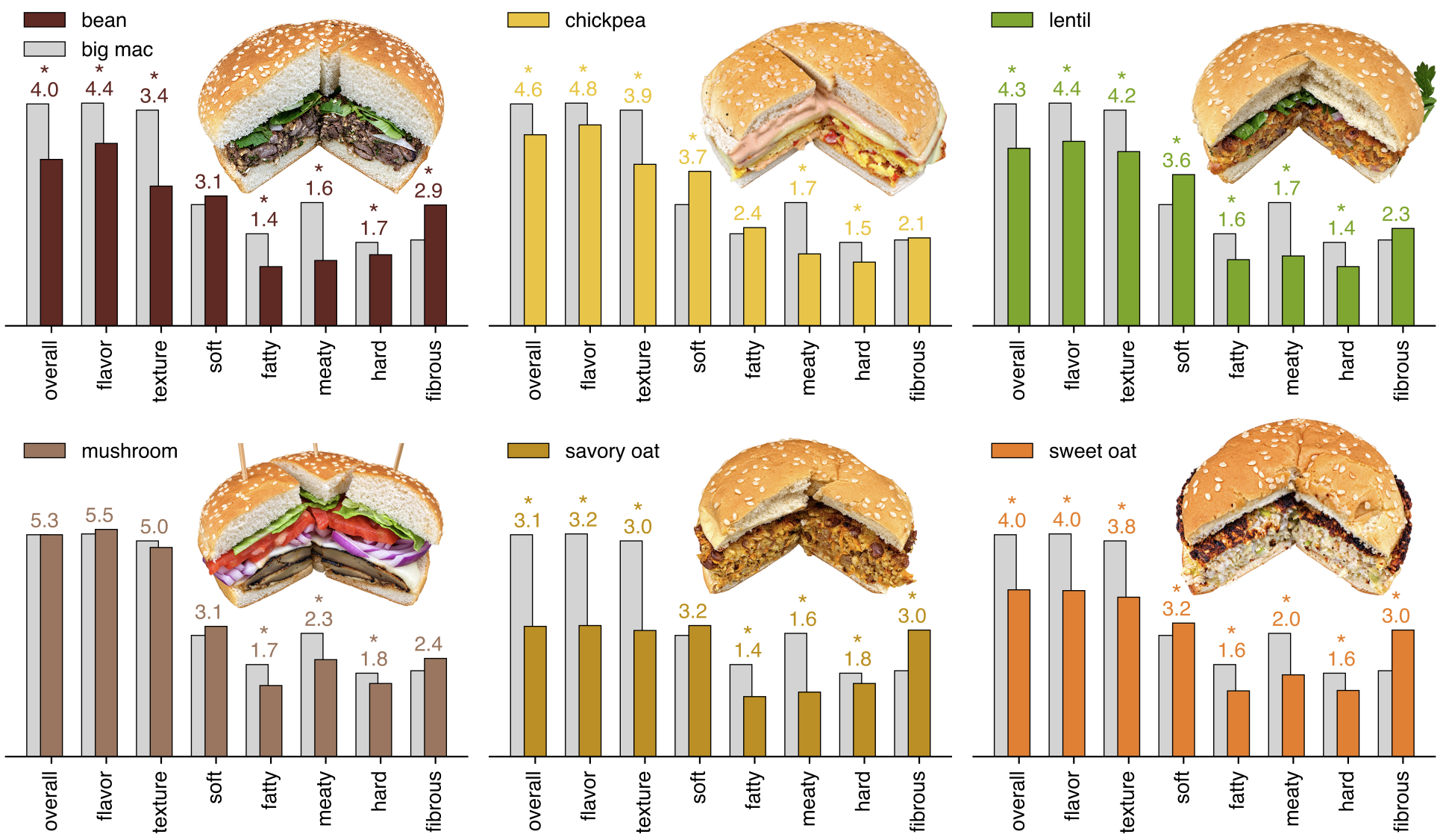}
\caption{{\sffamily{\textbf{Sensory scores for the six plant-based burgers compared to the Big Mac.}}} Scores of overall liking, flavor, and texture rated on a seven-point Likert scale, and softness, fattiness, meatiness, hardness, and fibrousness, rated on a five-point Likert scale. The mushroom burger has mean scores similar to the Big Mac for overall liking, flavor, and texture.}
\label{fig1}
\end{figure*}

Figure \ref{fig1} shows the mean consumer scores for the six plant-based burgers compared to the Big Mac \cite{tacGenerativeArtificialIntelligence2026}. Respondents rated overall, flavor, and texture liking on a seven-point Likert scale and five textural attributes, softness, fattiness, meatiness, hardness, and fibrousness, on a five-point Likert scale. The mushroom burger has mean scores similar to the Big Mac in overall liking (5.3 vs.~5.3), flavor (5.5 vs.~5.4), and texture (5.0 vs.~5.2). The other plant-based burgers receive lower liking scores, where the chickpea and lentil burgers perform best among this group (4.6 and 4.3).
The textural ratings also distinguish the Big Mac from the plant-based burgers. Respondents rate the Big Mac as meatier (3.0\,$\pm$\,1.1) and harder (2.0\,$\pm$\,0.9) than all six AI burgers and as less soft (2.9\,$\pm$\,1.0) and fibrous (2.1$\pm$0.9). The Big Mac receives the second-highest fattiness score after the chickpea burger. Notably, the sweet oat burger outperforms the savory oat burger by almost one point across all three liking scores, overall liking (4.0 vs.~3.1), flavor (4.0 vs.~3.2), and texture (3.8 vs.~3.0).

\begin{figure*}[t!]
\centering
\includegraphics[width=.9\textwidth]{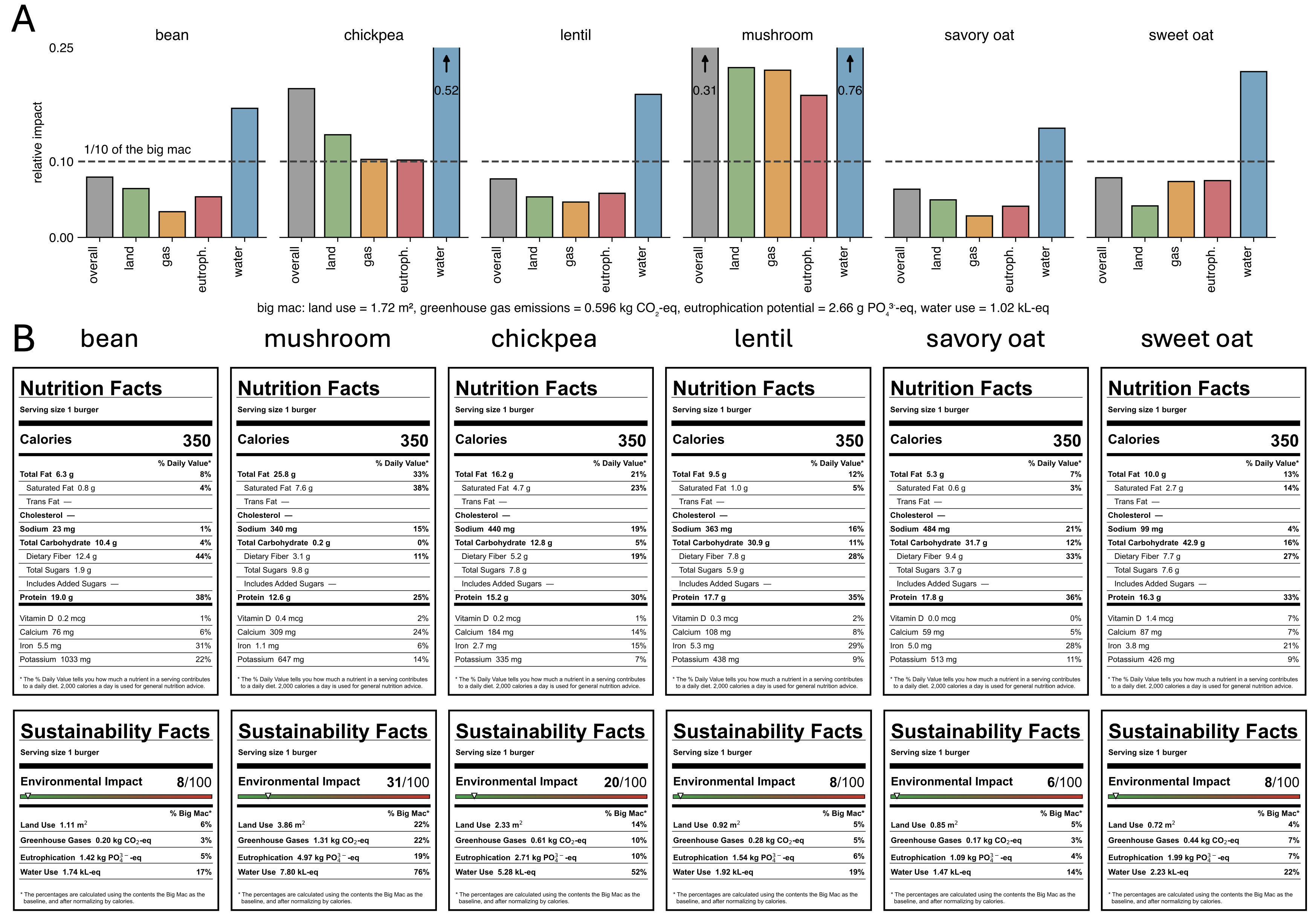}
\caption{{\sffamily{\textbf{Environmental impact scores for the six plant-based burgers compared to the Big Mac.}}} Scores of overall impact, land use, greenhouse gas emissions, eutrophication potential, and water use (top) and proposed sustainability facts label compared to the standard nutrition facts label for the plant-based burgers (bottom).}
\label{fig2}
\end{figure*}

Figure \ref{fig2} quantifies the environmental impact of each recipe in terms of land use, greenhouse gas emissions, eutrophication potential, and scarcity-weighted water use associated with ingredient production. We obtain these measures from life-cycle assessment data \cite{pooreReducingFoodsEnvironmental2018}, combine them into an overall Environmental Impact Score through a weighted average \cite{clarkEstimatingEnvironmentalImpacts2022}, and compare the four impact categories and the overall score for each plant-based burger with those of the Big Mac (Fig. \ref{fig2}, top). Most plant-based burgers produce overall impacts near one-tenth of the Big Mac value. The mushroom burger produces the largest impact among the plant-based formulations, but its overall score remains below one-third of the Big Mac value. Across all four impact categories, water use shows the smallest reduction compared to the Big Mac (33.3\,$\pm$\,25.1\%), but remains 17\% to 76\% below the Big Mac value for every plant-based burger. Thus, the mushroom burger achieves mean liking scores similar to those of the Big Mac in consumer liking while producing substantially lower environmental impacts.
Environmental impact spans several dimensions, which complicates communication with consumers. To present this information in a familiar and accessible format, we propose a Sustainability Facts Label modeled after the standard Nutrition Facts Label (Fig. \ref{fig2}, bottom). The label reports land use, greenhouse gas emissions, eutrophication potential, and water use as both absolute values and percentages of the corresponding Big Mac values. It also highlights the overall Environmental Impact Score with a larger font and a color scale for rapid interpretation. 

\begin{figure*}[t!]
\centering
\includegraphics[width=.9\textwidth]{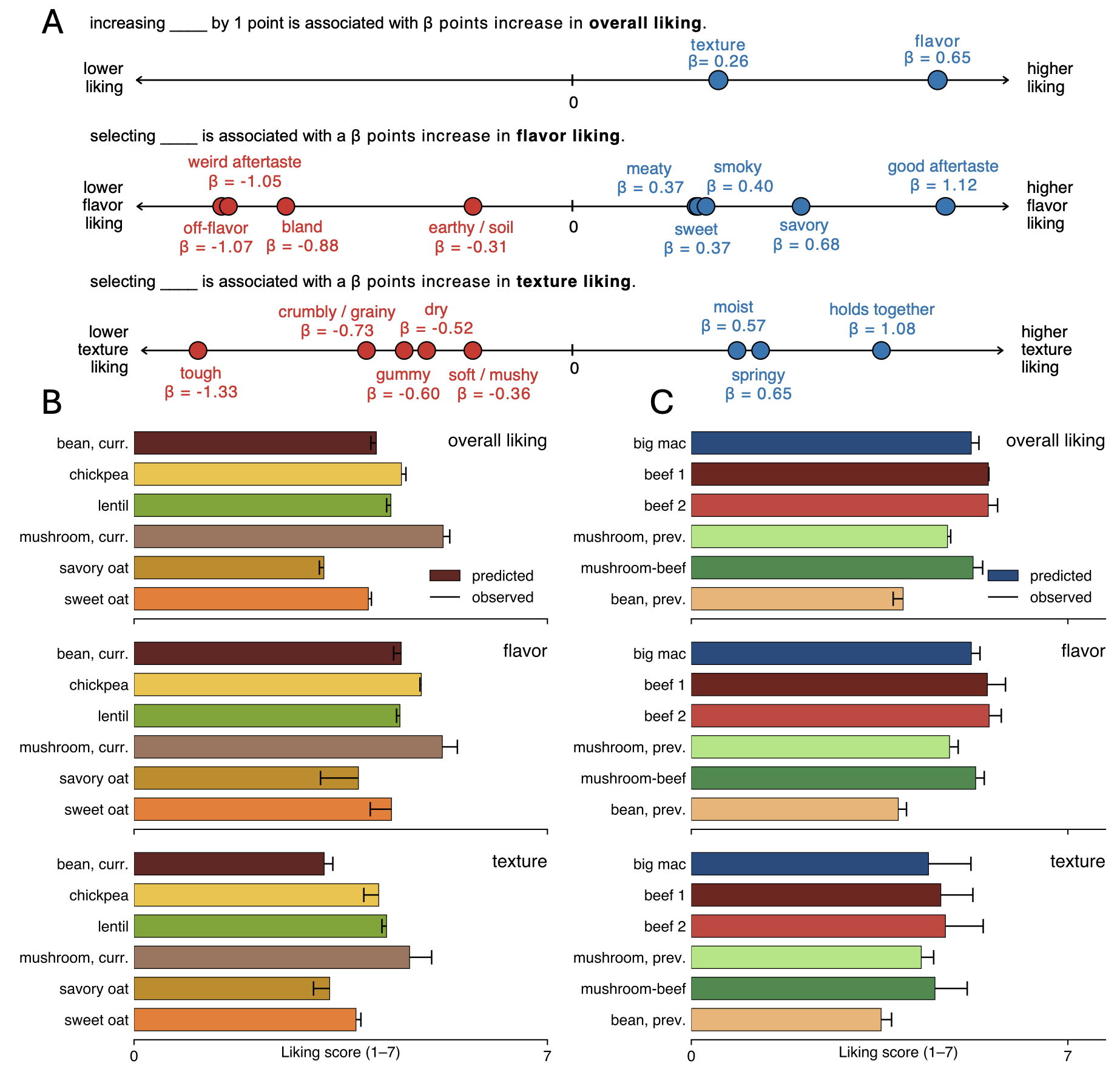}
\caption{{\sffamily{\textbf{Mixed-effects modeling of overall liking, flavor, and texture.}}} Mixed effects coefficients for models of overall liking, flavor liking, and texture liking (top). Predicted and observed scores using the constructed mixed-effects models for the 6 burgers in the current survey (bottom, left). Validation of the model with unseen data from the previous survey (bottom, right).}
\label{fig3}
\end{figure*}

Figure \ref{fig3} illustrates the sensory attributes that contribute to overall liking, flavor, and texture from mixed effect modeling. 
The mixed-effects model for overall liking uses flavor and texture liking as predictors;
the flavor and texture models use the corresponding check-all-that-apply (CATA) attributes as predictors.
Flavor and texture liking are positively associated with overall liking (top), although flavor has the larger coefficient ($\beta=0.65$ versus $\beta=0.26$). These values closely match the coefficients from our previous study ($\beta=0.65$ and $\beta=0.28$) \cite{tacTextureIndependentlyDrives2026}.
Among the flavor attributes, good aftertaste has the largest positive coefficient ($\beta=1.12$). With all other predictors held constant, selecting this attribute corresponds to a 1.12-point increase in flavor liking on the seven-point scale. In contrast, off-flavor ($\beta=-1.07$) and weird aftertaste ($\beta=-1.05$) have the largest negative flavor coefficients. Among the texture attributes, holds together has the largest positive coefficient ($\beta=1.08$), whereas tough has the largest negative coefficient ($\beta=-1.33$).
For the six burgers in the current survey, the mixed-effects models produce mean absolute errors of 0.080, 0.244, and 0.201 points for overall liking, flavor, and texture (bottom left). When applied to independent survey data from our previous study, the models produce corresponding errors of 0.125, 0.200 , and 0.519 points (bottom right).

\begin{figure*}[t!]
\centering
\includegraphics[width=.9\textwidth]{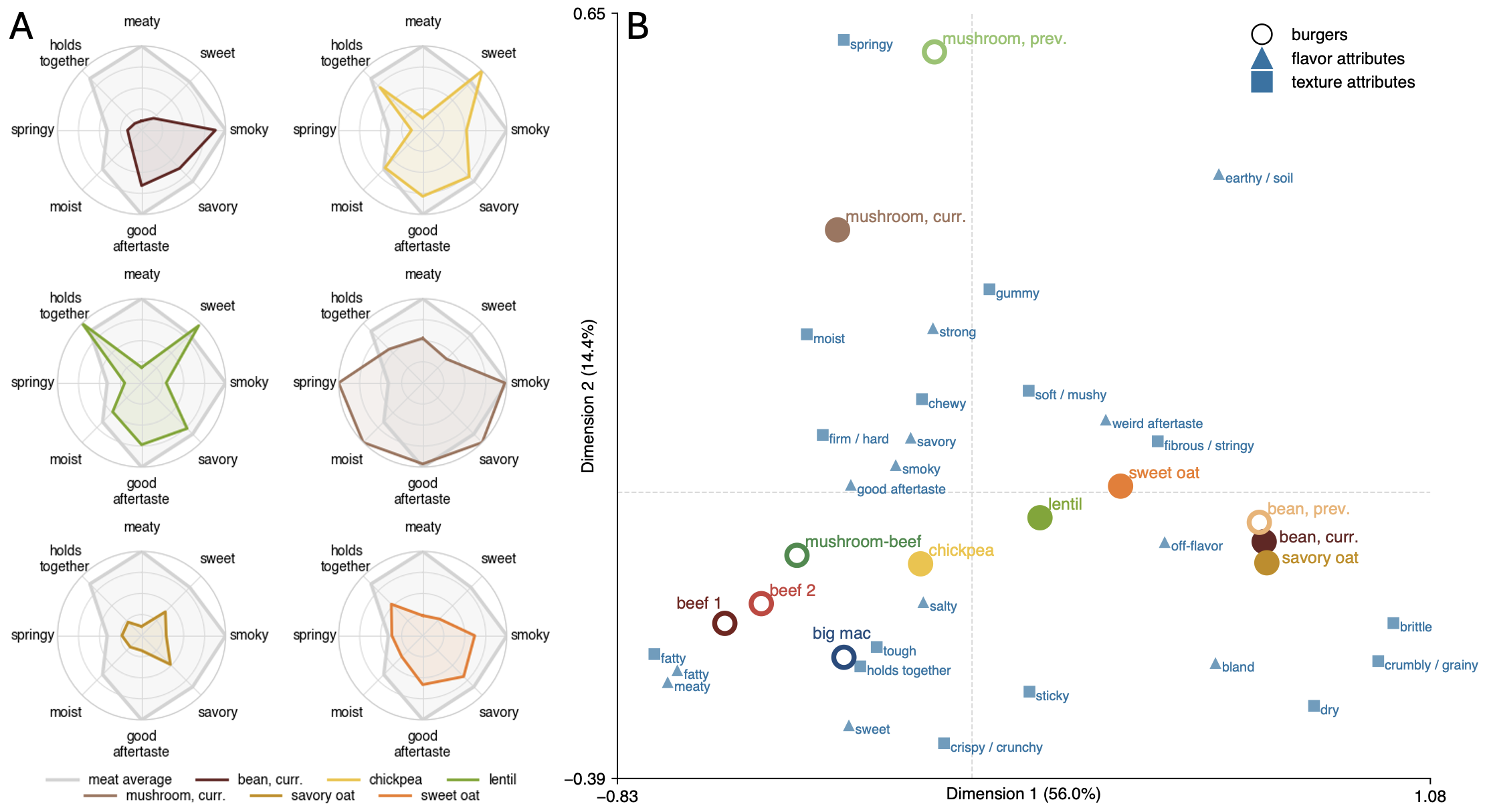}
\caption{{\sffamily{\textbf{Spider plots and correspondence analysis biplot for the six plant-based burgers compared to four meat-containing burgers.}}} 
The spider plots compare positive flavor- and texture-related attributes of all six plant-based burgers against the average of all four meat-based burgers from a previous study. 
The correspondence analysis biplot compares the sensory profiles of all twelve burgers, where filled circles denote current formulations and open circles denote formulations from the previous study. 
The four meat-based burgers cluster in the lower-left region, while the plant-based burgers span a broader sensory space. 
The previous and current bean burgers are nearly co-located, while the previous and current mushroom burgers occupy the same upper-left region but remain distinct.}
\label{fig4}
\end{figure*}

Figure \ref{fig4} compares the flavor- and texture-related attributes of the six current plant-based burgers with six burgers from our previous study \cite{tacGenerativeArtificialIntelligence2026}. The spider plots compare 
eight flavor- and texture-related attributes with positive mixed-effects coefficients,$\beta>0$, for all six plant-based burgers against the average of all four meat-containing burgers from a previous study (left). Across these attributes, the current mushroom burger exhibits the sensory profile most similar to the meat-burger average. The correspondence-analysis biplot expands the comparison to all six current and six previous burgers and all measured flavor and texture attributes (right). The first two dimensions explain 70.4\% of the total inertia. The four meat-containing burgers, the Big Mac, beef 1, beef 2, and mushroom-beef, cluster in the lower-left region, whereas the plant-based burgers span a broader sensory space. Notably, the previous and current bean burgers, which contain the same ingredients in different quantities, are nearly co-located. The two independently formulated mushroom burgers occupy the same upper-left region but remain distinct. Together, these patterns suggest that primary ingredient identity is associated with the broad sensory profile, while differences in formulation are associated with differences among individual products within this sensory space.
\section*{\sffamily{\bfseries{Discussion}}}
Our results show that generative AI can guide the design of plant-based burgers that combine consumer appeal with substantially lower environmental impacts. Among the six formulations, the mushroom burger achieves mean overall, flavor, and texture liking scores most similar to the Big Mac. Across the eight sensory attributes positively associated with liking, it also exhibits the profile most similar to the meat-burger average. At the same time, its overall environmental impact remains below one-third of the Big Mac value. The other plant-based burgers produce even lower impacts, often near one-tenth of the Big Mac value, but they also receive lower sensory scores. These findings illustrate the central challenge in sustainable food design: minimizing environmental impact alone does not maximize consumer acceptance.


The correspondence analysis provides a broader view of the sensory design space. The four meat-containing burgers cluster together, whereas the plant-based burgers span a wider range of sensory profiles. The previous and current bean burgers are nearly co-located despite their different ingredient quantities, while the two independently formulated mushroom burgers occupy the same broad region but remain distinct. These patterns suggest that primary ingredient identity organizes the broad sensory profile, while ingredient quantities and formulation differentiate individual products within this space. Because the products were evaluated in separate surveys, these cross-study comparisons remain descriptive.

The mixed-effects models quantify how flavor and texture relate to consumer preference. Flavor has a stronger relationship with overall liking than texture, 
although both are positively associated with overall liking. This result closely agrees with our previous study \cite{tacTextureIndependentlyDrives2026} and with sensory research that identifies flavor and texture as major determinants of plant-based meat acceptance \cite{vandenbedem2026}. The attribute-level models provide more specific design guidance. 
For flavor, good aftertaste has the largest positive association with liking, while off-flavor and weird aftertaste have the strongest negative associations.
For texture, holds together has the strongest positive association with liking, while tough has the strongest negative association.
These coefficients translate consumer responses into quantitative objectives that can guide subsequent recipe generation and refinement.

The models also retain predictive accuracy when applied to independent data from our previous survey. This agreement suggests that the relationships among sensory attributes, flavor, texture, and overall liking may extend beyond the formulations used to train the models. The larger validation error for texture indicates greater variation across products or consumer groups and identifies texture as an important target for additional data collection. Future generative models could incorporate these learned relationships directly into their objective functions and prioritize recipes that promote desirable attributes while avoiding off-flavors, unpleasant aftertastes, and toughness.

The environmental analysis demonstrates the magnitude of the potential benefit. Every plant-based burger has lower estimated values for the four environmental impact categories relative to the Big Mac, although the size of the reduction varies by formulation and category. These results agree with prior life-cycle assessments that identify beef as one of the most resource-intensive common foods \cite{xu2021}. They also show why sustainable product design requires several environmental measures \cite{pooreReducingFoodsEnvironmental2018}. A recipe may reduce greenhouse gas emissions and land use substantially, while producing a smaller reduction in water use. The proposed sustainability facts label summarizes these trade-offs in a format that consumers already recognize. However, the present study introduces the label as a communication framework and does not yet test whether consumers understand it or whether it changes their choices.

Several limitations define the scope of these conclusions. First, the study evaluates six AI-generated, chef-refined plant-based burgers in one restaurant context, so the results may not generalize to other recipes, preparation methods, or consumer populations. Because the chef refines the ingredient quantities based on culinary expertise, the sensory outcomes evaluate the complete human--AI workflow rather than autonomous AI generation. Second, although the Big Mac scores originate from the same restaurant and similar population demographics, a direct, randomized, and blinded comparison within the same participant cohort would provide a stronger test of sensory equivalence. Third, the life-cycle estimates depend on ingredient-level databases, weighting choices, geographic assumptions, and system boundaries, and do not capture every consequence of production, preparation, packaging, transportation, or food waste. Finally, the current analysis focuses on liking and environmental impact and does not establish nutritional superiority, affordability, commercial feasibility, or long-term consumer adoption.

Despite these limitations, the study establishes a human-centered, quantitative workflow that connects generative food design, consumer sensory evaluation, environmental assessment, and statistical models of preference. The results identify a plant-based mushroom burger that approaches an iconic meat benchmark in consumer appeal while producing substantially lower environmental impacts. They also reveal which flavor and texture attributes most strongly predict liking. Together, generative models propose formulations, culinary expertise translates them into foods, and sensory models learn from consumer responses to guide future designs toward products that people want to eat.

\small
\section*{\sffamily{\bfseries{Methods}}}\label{methods}
We previously developed and experimentally validated a generative artificial intelligence platform for food design \cite{tacGenerativeArtificialIntelligence2026}. The original training dataset contained 2,216 recipes from food.com but relatively few plant-based formulations, which limited the model's representation of plant-based ingredient combinations. In this study, we augment the training dataset with 469 human-designed plant-based burger recipes from cooks.com. We retrain the generative model on this combined dataset to extend the framework to the design of plant-based burgers.\\[6.pt]
{\sffamily{\bfseries{Generative model.}}} 
We represent each burger recipe as a hybrid discrete--continuous object with two sequential tasks: ingredient selection and ingredient quantification. Our generative AI architecture assigns these tasks to a multinomial diffusion model and a Gaussian diffusion model \cite{tacGenerativeAIMaterial2026a}. We refer to these models as the mask model and weight model.
The mask model generates a recipe mask, a binary vector $\mathbf{m}$, where each entry $m_i$ encodes the presence or absence of ingredient $i$.
The multinomial diffusion process progressively corrupts and reconstructs these binary recipe masks through categorical probability distributions \cite{NEURIPS2021_67d96d45}.
The weight model uses each generated mask to condition the ingredient quantification process. This score-based generative model applies a Gaussian diffusion process to the ingredient-weight vectors \cite{songScoreBasedGenerativeModeling2021}. The model learns the score, 
$\nabla_{\mathbf{w}} \log p_t(\mathbf{w}\mid\mathbf{m})$, of the conditional probability distribution of weight vectors at each diffusion time. A feed-forward neural network approximates the score function as a function of the recipe mask and diffusion time.\\[6.pt]
{\sffamily{\bfseries{Human--AI recipe development.}}} 
An executive chef translates the generated formulations into standardized preparation protocols and refines the proposed ingredient quantities based on culinary expertise while preserving the AI-generated ingredient selections. The Supplementary Information compares the AI-generated and chef-refined quantities and provides the complete recipes and preparation protocols.\\[6.pt]
{\sffamily{\bfseries{Sensory survey.}}} 
We conduct a blinded sensory survey with $n=101$ community participants in an operating restaurant in San Francisco. We recruit participants by email. We exclude individuals who are younger than 18 years, follow vegan or vegetarian diets, consume burgers fewer than four times per year, or report allergies to dairy, wheat, soy, eggs, corn, or alliums.
We prepare all burgers in the restaurant kitchen according to standardized recipe instructions and serve them fresh. We do not disclose the burger identities or ingredients. Each participant evaluates all six burgers in a randomized order and uses water to rinse the palate between consecutive samples.
Participants evaluate overall liking, flavor, and texture on a seven-point Likert scale \cite{st.pierreMechanicalSensorySignature2024} and answer check-all-that-apply (CATA) questions about flavor- and texture-related attributes \cite{vandenbedem2026}. Participants also evaluate five sensory attributes of the burger patty on a five-point Likert scale: softness, hardness, fattiness, meatiness, and fibrousness. The Stanford University Institutional Review Board determined that this project does not meet the definition of human-subjects research under 45 CFR 46.102.\\[6.pt]
{\sffamily{\bfseries{Environmental impact and sustainability facts.}}} 
We quantify land use, greenhouse gas emissions, eutrophication potential, and scarcity-weighted water use using ingredient-level life-cycle assessment data \cite{pooreReducingFoodsEnvironmental2018}. For each burger, we multiply the quantity of each ingredient by its corresponding impact factor, sum the ingredient-level contributions \cite{clarkEstimatingEnvironmentalImpacts2022}, and  combine the four impact categories into a single environmental impact score \cite{tacGenerativeArtificialIntelligence2026}. We report the individual impact categories and overall score on a sustainability facts label, both as absolute values per burger and relative to the Big Mac.
\\[6.pt]
{\sffamily{\bfseries{Correspondence analysis.}}} 
We use correspondence analysis \cite{abdiCorrespondenceAnalysis2017} to examine relationships among the 12 burgers from the present and previous studies and the flavor- and texture-related attributes collected through the check-all-that-apply (CATA) questions. We construct a contingency table $\mathbf{N}=(n_{ij})$, where rows represent burgers, columns represent flavor and texture CATA attributes, and each entry $n_{ij}$ gives the number of participants who select attribute $j$ for burger $i$. We obtain the relative-frequency matrix $\mathbf{P}=\mathbf{N}/n$, where $n$ is the total number of selections across all attributes and burgers. We compute the row and column marginal proportions as $r_i=\sum_j p_{ij}$ and $c_j=\sum_i p_{ij}$ and center the data with respect to the independence model,
$\mathbf{Z}
=\mathbf{D}_{\rm r}{}^{-1/2}
[\,\mathbf{P}-\mathbf{r}\,\mathbf{c}^{\rm t}\,]\,
 \mathbf{D}_{\rm c}{}^{-1/2}$,
where $\mathbf{D}_{\rm r}$ and $\mathbf{D}_{\rm c}$ denote diagonal matrices of row and column masses. We apply a singular value decomposition to $\mathbf{Z}$ to obtain the principal axes and corresponding singular values and define the eigenvalues as $\lambda_k=\sigma_k^2$ to quantify the inertia explained by each dimension.
For the spider plots, we compare the selection frequencies of positively associated flavor and texture attributes for each plant-based burger with their mean frequencies across the four meat-containing burgers from the previous study.\\[6.pt]
{\sffamily{\bfseries{Mixed-effects models.}}} 
We fit linear mixed-effects models to identify sensory measures associated with overall liking, flavor, and texture while accounting for repeated measurements within participants \cite{gelman_hill_2006}. The first model treats overall liking as the dependent variable and flavor and texture as fixed-effect predictors. The next two models treat flavor and texture as dependent variables and include the relevant binary CATA attributes as fixed-effect predictors. All three models include a random intercept for each participant to account for individual baseline differences. We fit the models to the data from the present survey and evaluate their predictions against the independent data from our previous survey without refitting. We quantify prediction error as the absolute difference between the
observed and predicted mean liking score for each burger and report the mean absolute error across burgers. \\[10.pt]
{\sffamily{\bfseries{Acknowledgements.}}}
We thank Executive Chef Justin Schneider 
for translating the generated formulations 
into preparation protocols and refining the ingredient quantities, 
Caroline Cotto from NECTAR at Food System Innovations 
for stimulating discussions, and 
Alice Wistar and Alex Weissman from Palate Insights 
for conducting the consumer survey.
We acknowledge 
access to the Stanford Marlowe Computing Platform.
This research was supported 
by the Schmidt Science Fellowship, 
in partnership with the Rhodes Trust, to Vahidullah Tac, and 
by the Stanford Doerr School of Sustainability Accelerator, 
Stanford Bio-X Snack Grant,
Bezos Earth Fund, 
NSF CMMI grant 2320933, and 
ERC Advanced Grant 101141626 to Ellen Kuhl.

\begin{thebibliography}{10}

\bibitem{keesing2022}
F Keesing, Diet for a small footprint.
\newblock {\emph{Proceedings of the National Academy of
  Sciences}} \textbf{119}, e2204241119 (2022).

\bibitem{tilmanGlobalDietsLink2014}
D Tilman, M Clark, Global diets link environmental sustainability and human
  health.
\newblock {\emph{Nature}} \textbf{515}, 518--522 (2014).

\bibitem{eshelLandIrrigationWater2014}
G Eshel, A Shepon, T Makov, R Milo, Land, irrigation water, greenhouse gas, and
  reactive nitrogen burdens of meat, eggs, and dairy production in the {{United
  States}}.
\newblock {\emph{Proceedings of the National Academy of
  Sciences}} \textbf{111}, 11996--12001 (2014).

\bibitem{pooreReducingFoodsEnvironmental2018}
J Poore, T Nemecek, Reducing food's environmental impacts through producers and
  consumers.
\newblock {\emph{Science}} \textbf{360}, 987--992 (2018).

\bibitem{shiRedMeatConsumption2023}
W Shi, X Huang, CM Schooling, JV Zhao, Red meat consumption, cardiovascular
  diseases, and diabetes: A systematic review and meta-analysis.
\newblock {\emph{European Heart Journal}} \textbf{44},
  2626--2635 (2023).

\bibitem{springmannAnalysisValuationHealth2016}
M Springmann, HCJ Godfray, M Rayner, P Scarborough, Analysis and valuation of
  the health and climate change cobenefits of dietary change.
\newblock {\emph{Proceedings of the National Academy of
  Sciences}} \textbf{113}, 4146--4151 (2016).

\bibitem{clarkMultipleHealthEnvironmental2019}
MA Clark, M Springmann, J Hill, D Tilman, Multiple health and environmental
  impacts of foods.
\newblock {\emph{Proceedings of the National Academy of
  Sciences}} \textbf{116}, 23357--23362 (2019).

\bibitem{crimarcoRandomizedCrossoverTrial2020}
A Crimarco, et~al., A randomized crossover trial on the effect of plant-based
  compared with animal-based meat on trimethylamine-{{N-oxide}} and
  cardiovascular disease risk factors in generally healthy adults: {{Study With
  Appetizing Plantfood}}---{{Meat Eating Alternative Trial}} ({{SWAP-MEAT}}).
\newblock {\emph{The American Journal of Clinical
  Nutrition}} \textbf{112}, 1188--1199 (2020).

\bibitem{szenderakConsumerAcceptancePlantBased2022}
J Szender{\'a}k, D Fr{\'o}na, M R{\'a}kos, Consumer {{Acceptance}} of
  {{Plant-Based Meat Substitutes}}: {{A Narrative Review}}.
\newblock {\emph{Foods}} \textbf{11}, 1274 (2022).

\bibitem{giezenaarProfilingNovelPlantbased2024}
C Giezenaar, et~al., Profiling the novel plant-based meat alternative category:
  {{Consumer}} affective and sensory response in the context of perceived
  similarity to meat.
\newblock {\emph{Food Research International}} \textbf{188},
  114465 (2024).

\bibitem{st.pierreMeatballMatchupPlant2026}
SR St.~Pierre, AO Koosis, N Zhang, E Kuhl, The meatball matchup: {{Plant}} vs.
  animal proteins on campus.
\newblock {\emph{Food Research International}} \textbf{240},
  119631 (2026).

\bibitem{datta2026}
B Datta, et~al., Artificial intelligence for food innovation.
\newblock {\emph{Nature Food}} \textbf{7}, 644--654 (2026).

\bibitem{tacGenerativeArtificialIntelligence2026}
V Tac, CD Gardner, E Kuhl, Generative artificial intelligence creates
  delicious, sustainable, and nutritious burgers.
\newblock {\emph{npj Science of Food}} \textbf{10}, 199
  (2026).

\bibitem{kuhlAIFoodAccelerating2025}
E Kuhl, {{AI}} for food: Accelerating and democratizing discovery and
  innovation.
\newblock {\emph{npj Science of Food}} \textbf{9}, 82
  (2025).

\bibitem{zhangFoodProductDesign2019}
X Zhang, T Zhou, L Zhang, KY Fung, KM Ng, Food {{Product Design}}: {{A Hybrid
  Machine Learning}} and {{Mechanistic Modeling Approach}}.
\newblock {\emph{Industrial \& Engineering Chemistry
  Research}} \textbf{58}, 16743--16752 (2019).

\bibitem{tacGenerativeAIMaterial2026a}
V Tac, E Kuhl, Generative {{AI}} for material design: {{A}} mechanics
  perspective from burgers to matter.
\newblock {\emph{Computer Methods in Applied Mechanics and
  Engineering}} \textbf{461}, 119171 (2026).

\bibitem{clarkEstimatingEnvironmentalImpacts2022}
M Clark, et~al., Estimating the environmental impacts of 57,000 food products.
\newblock {\emph{Proceedings of the National Academy of
  Sciences}} \textbf{119}, e2120584119 (2022).

\bibitem{tacTextureIndependentlyDrives2026}
V Tac, AO Koosis, E Kuhl, Texture independently drives liking in
  {{AI-generated}} alternative protein burgers.
\newblock {\emph{Foods}} \textbf{15}, 2026 (2026).

\bibitem{vandenbedem2026}
SD van~den Bedem, E Kuhl, C Cotto, Open-source benchmarking of plant-based and
  animal meats.
\newblock {\emph{Foods}} \textbf{15}, 2112 (2026).

\bibitem{xu2021}
X Xu, et~al., Global greenhouse gas emissions from animal-based foods are twice
  those of plant-based foods.
\newblock {\emph{Nature Food}} \textbf{2}, 724--732 (2021).

\bibitem{NEURIPS2021_67d96d45}
E Hoogeboom, D Nielsen, P Jaini, P Forr{\'e}, M Welling, Argmax flows and
  multinomial diffusion: {{Learning}} categorical distributions in {\em
  Advances in Neural Information Processing Systems}, eds.{} M Ranzato, A
  Beygelzimer, Y Dauphin, P Liang, JW Vaughan.
\newblock (Curran Associates, Inc.), Vol.{}~34, pp. 12454--12465 (2021).

\bibitem{songScoreBasedGenerativeModeling2021}
Y Song, et~al., Score-{{Based Generative Modeling}} through {{Stochastic
  Differential Equations}} (2021).

\bibitem{st.pierreMechanicalSensorySignature2024}
SR St.~Pierre, et~al., The mechanical and sensory signature of plant-based and
  animal meat.
\newblock {\emph{npj Science of Food}} \textbf{8}, 94
  (2024).

\bibitem{abdiCorrespondenceAnalysis2017}
H Abdi, M B{\'e}ra, Correspondence {{Analysis}} in {\em Encyclopedia of
  {{Social Network Analysis}} and {{Mining}}}, eds.{} R Alhajj, J Rokne.
\newblock (Springer New York, New York, NY), pp. 1--12 (2017).

\bibitem{gelman_hill_2006}
A Gelman, J Hill, {\em Data Analysis Using Regression and
  Multilevel/Hierarchical Models}, Analytical Methods for Social Research.
\newblock (Cambridge University Press, Cambridge), (2006).
\end{thebibliography}

\clearpage


\section*{Supplementary material}

# Generative AI enables human-centered design of sustainable plant-based foods

## Supplementary Information

This Supplementary Information documents how generative AI and culinary expertise jointly produce the six plant-based burgers evaluated in this study. It presents the guidelines provided to the executive chef, compares AI-generated and chef-refined ingredient quantities, and provides standardized preparation protocols and photographic records for the bean, chickpea, lentil, mushroom, savory oat, and sweet oat burgers. Additional photographs document food preparation and presentation during the restaurant survey. These materials support reproducibility and clarify the human contribution to the generative food-design framework.

# Guidelines for the Executive Chef

This document is intended to be used by the Executive Chefs. It presents the ingredient lists and their corresponding quantities and describes how the lists should be interpreted.

1. Each one of the listed recipes amounts to 350 calories exactly. The chefs should use the **given weights of the ingredients without any changes**.
2. The recipes don't list buns. The chefs should choose **1 standard sesame seed bun for all 6 burgers**.
3. There are two separate recipes given under **"oat and fruit burger"**, one uses apples and the other uses mangoes. The chefs are asked to **choose one** that they think will be most preferred by consumers, but not mix the two recipes.
4. The ingredient lists only show the ingredients, but all processing steps are to be decided by the chefs.
5. The bean burger is a repeat of the bean burger from the previous study. The chefs are asked to design an **alternative way of preparing this burger**.
6. The previous study also contained a mushroom burger, where the chefs decided to slice the mushroom into pieces. We ask that the chefs **grind/chop the mushroom into a patty this time**.
7. The chefs should decide which parts of the ingredient lists should be used in the patty, and which parts should be layered on top. This even includes dividing some ingredients into two or more parts. For example, if the recipe mentions "onions", the chef could decide to mince some of it and add to the patty, and slice some for layering on top of the patty. In such cases, we ask the chefs to record the weights of the parts.
8. The weights are generally meant to be dry and uncooked weights. If the chef believes that the weight listed for an ingredient is **too high/low for it to be the dry/uncooked weight** (e.g. beans), please let us know for further consideration.
9. **Salt**: The ingredient lists don't show salt. The chefs should adjust salt to taste (and record the amount they used).
10. All of the listed ingredients are meant to be used in preparing the burgers, **but not all of them have to stay on the burger**. Examples:
    a. Vegetables and fruits have peels and stems that can be removed after weighing.
    b. If the recipe lists "oil", and if the chef decides that the oil is best used for frying/sauteing, then the oil may remain in the pan after sauteing.
    c. Some herbs like rosemary may be used to add aroma to the rest of the ingredients and disposed afterwards.

11. Some of the recipes may result in “overflowing” burgers, this is acceptable but the chefs are still kindly asked to try their best in preparing/cooking the burgers in a way that is presentable, and that the burger has a structural integrity.
12. The ingredient lists do not mention specific details like variety of the ingredients. The chefs are free to make their own decisions about this. Example:
    a. Vegetables can be any variety that the chef prefers (e.g. red onions, white onions, etc (but not green onions, which are a different product)).
13. “**Other cheese**” includes any cheese that is <u>**not**</u> one of these: blue cheese, cheddar cheese, cottage cheese, cream cheese, feta, goat cheese, gouda, monterey, monterey jack, mozzarella, parmesan, provolone. We ask the chef to select a cheese that they think best fits the rest of the ingredients.
14. “**Other sauce**” includes any sauce that is <u>**not**</u> one of these: barbecue sauce, buffalo sauce, cream, curry sauce, hoisin sauce, honey, hot pepper sauces, ketchup, mustard, pesto, ranch, relish, salsa, soy sauce, steak sauce, teriyaki sauce, vinegar, wasabi, Worcestershire sauce. We ask the chef to select a sauce that they think best fits the rest of the ingredients.
15. **We kindly ask the chefs to put their best creativity and artistic skills to work to turn these ingredient lists to the best burgers possible, and wholeheartedly thank them for their help in this study!**

# Bean Burger

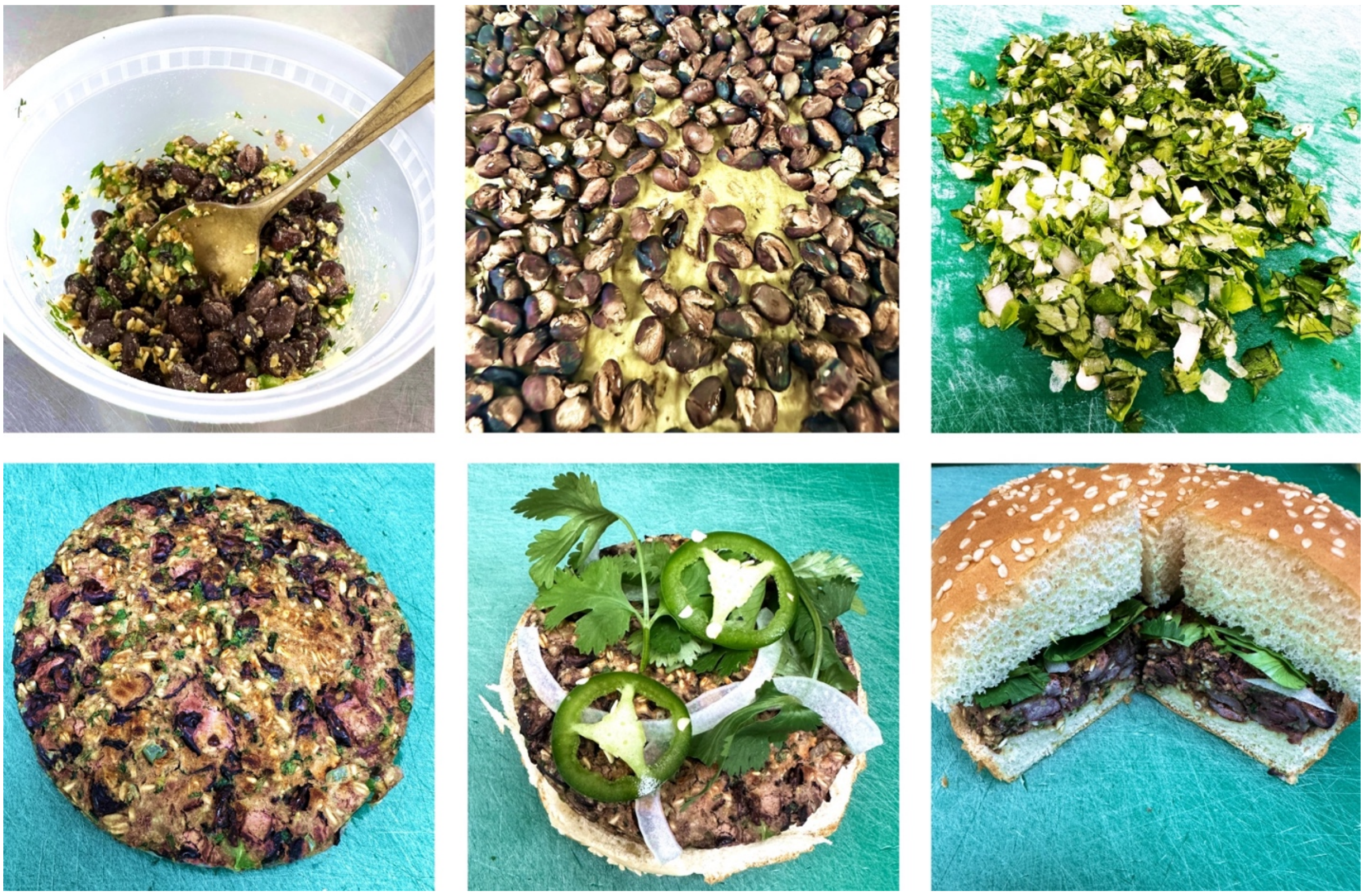

**Figure S1. Preparation of the bean burger.** Representative stages show ingredient preparation and mixing (top), the cooked bean patty, burger assembly, and the final burger in cross-section (bottom).

**<u>The Bean Burger</u>**

**Per burger:**

72 g canned black beans, rinsed & "dried"
16 g quick steel-cut oats
4 g finely diced onion
2 g finely diced jalapeño
2 g chopped cilantro
9 g whisked whole egg
3 g cornflour
4 g olive oil
1g salt

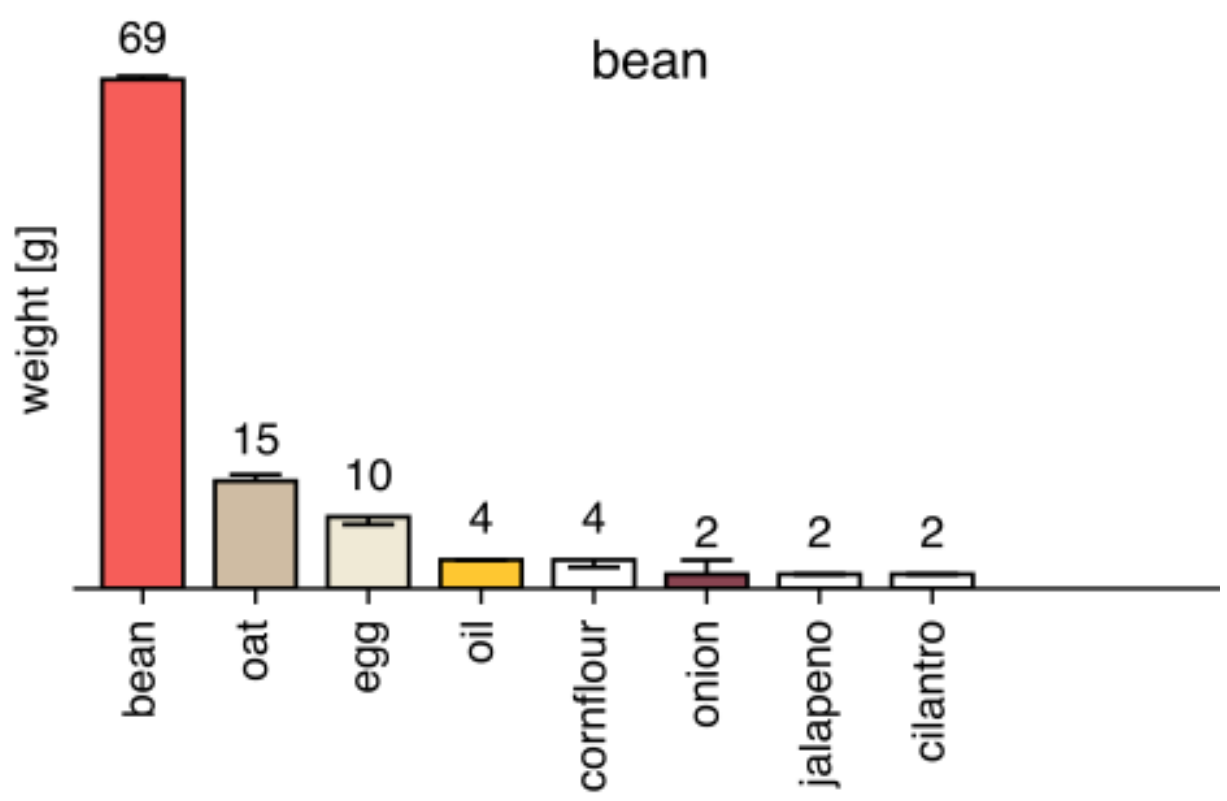


**Garnish:**
3 g finely sliced onion
2 g thinly sliced jalapeño
1 g fresh cilantro leaves & tender stems

**Prep:**
Spread the drained and rinsed beans on a lined sheet tray & place in a 350°f oven for 3-5 minutes to dry the immediate surface moisture.

Mix the beans, oats, onion, jalapeño, cilantro, egg, cornflour, oil, salt, and pepper until cohesive but not fully mashed. Cover and let rest for 10 minutes to hydrate the oats.

Shape into 30 x 113 g compact patties, pressing firmly, and chill.

Pre-weigh and measure garnish ingredients (3 g onion, 2 g sliced jalapeño, 1 g fresh cilantro) and put in a Ziplock bag, one for each burger (30 total).

**Day of:**

Cook the patty. Pan-sear over medium heat 4-5 minutes per side until browned and heated through (165°f)

Assemble the burger with a toasted sesame seed bun and the additional toppings.

# Chickpea Burger

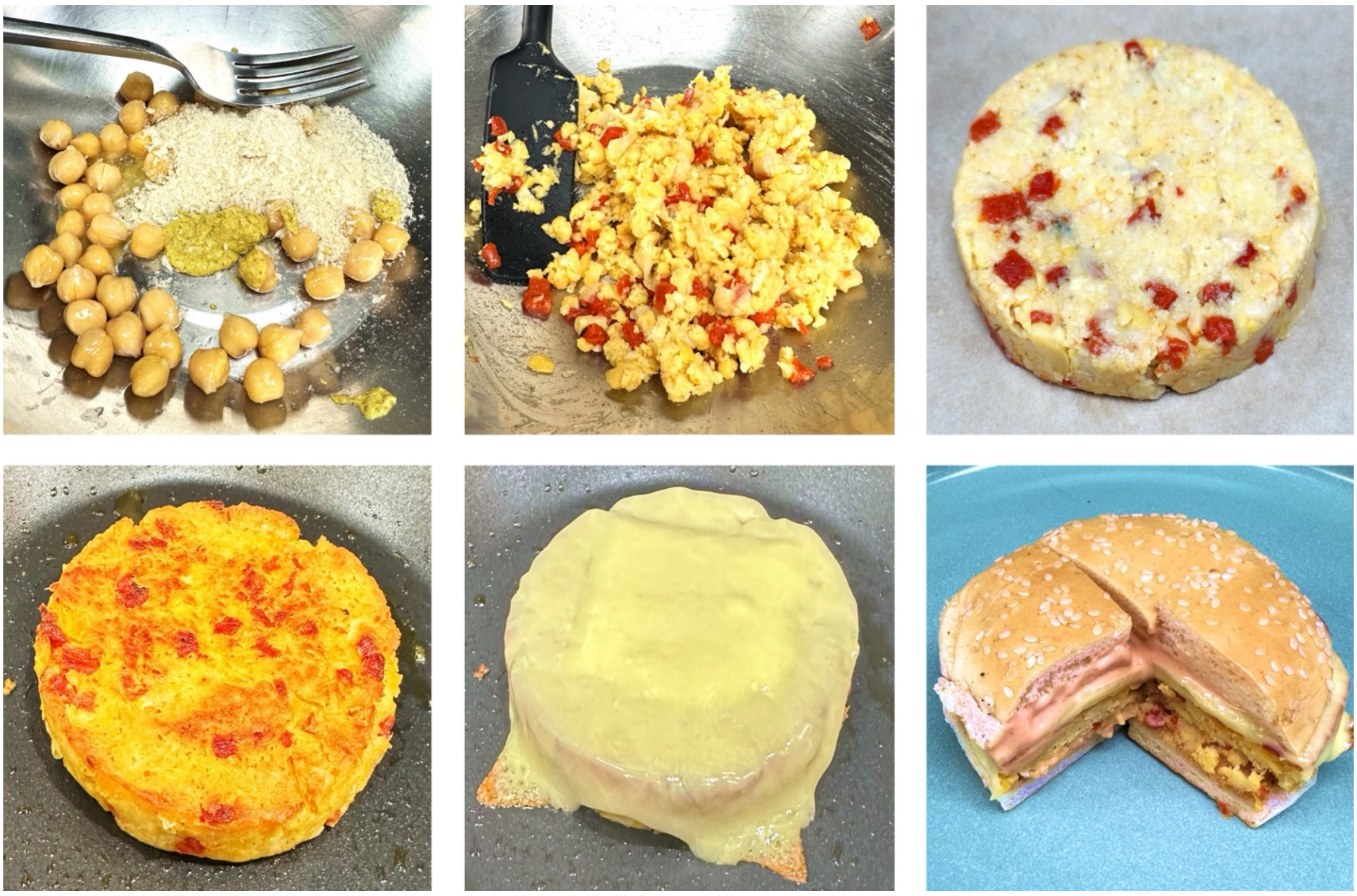

**Figure S2. Preparation of the chickpea burger.** Representative stages show ingredient preparation and mixing (top), the cooked bean patty, burger assembly, and the final burger in cross-section (bottom).

**<u>The Chickpea Burger</u>**

**Per burger:**

55g Chickpea, canned (drained & rinsed)
28g Cheddar Cheese, sliced
15g Mayo (Just, vegan brand)
11g Chilli, canned piquillo, small diced
10g Ketchup
8g Breadcrumbs, Panko
8g Egg, whisked
5g olive oil
4g Dijon Mustard
1g salt

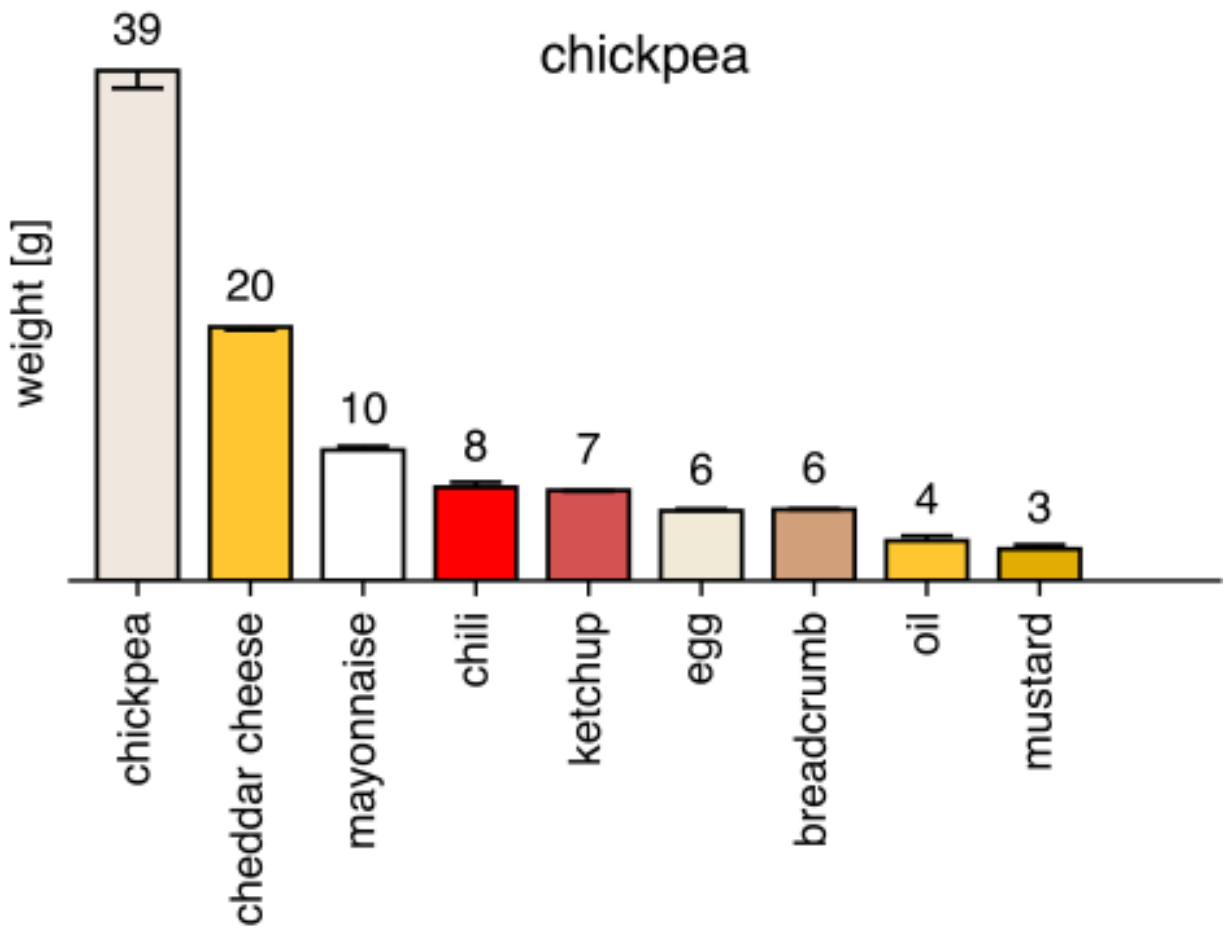


**Prep:**

Mash together the egg, breadcrumbs and chickpeas with a fork. Fold in the diced piquillo, mustard and salt. Form into 30 x 92 g patties.

Pre-measure cheddar cheese slices (28g each) and put into a labeled plastic bag, one for each burger (30 total bags).

Pre-measure ketchup (10g) and mayo (15g), mix, and put into a labeled condiment container, one for each burger (30 total containers).

**Day of:**

Cook in a non-stick pan set to 350°F using the remaining oil for 3.5 minutes on each side until golden brown, after flipping, add the cheese and allow to melt

Combine the ketchup and mayo to make a sauce and apply to a toasted sesame seed bun

# Lentil Burger

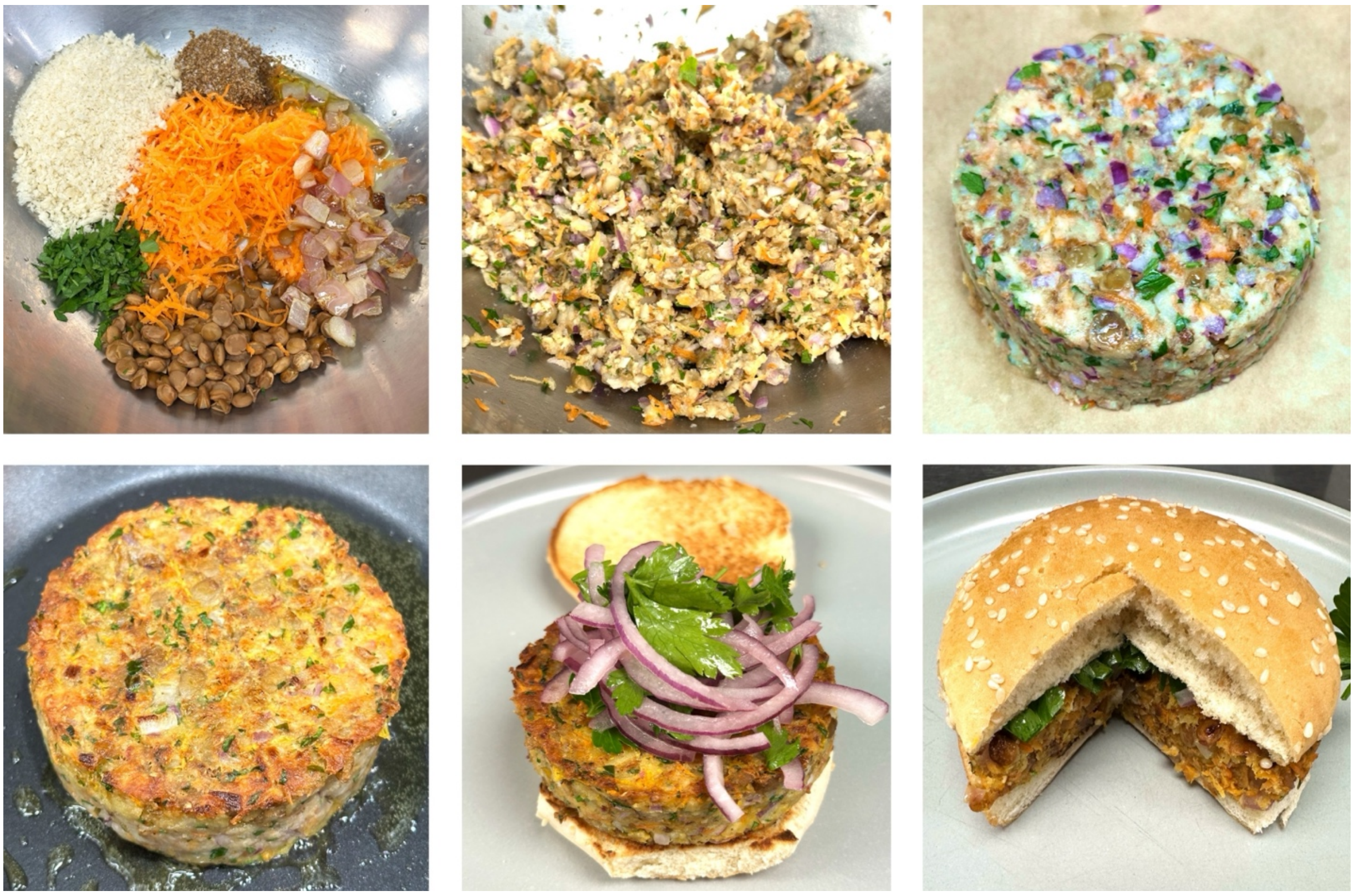

**Figure S3. Preparation of the lentil burger.** Representative stages show ingredient preparation and mixing (top), the cooked bean patty, burger assembly, and the final burger in cross-section (bottom).

**The Lentil Burger**

**Per burger:**

31.7 g Lentil, canned (drained,rinse, *dried*)
22.3 g Onion, red (dice 16.7 g patty, 5.6 g topping)
19.5 g carrot, finely shredded
12.8 g Egg, whisked
10 g Breadcrumbs, Panko
6.1 g Lemon, juice
4.5 g Olive oil
2.2 g Parsley, leaves (1.1 g patty, 1.1 g topping)
1.7 g AP Flour
1.7 g Coriander Seed, ground
.9 g salt

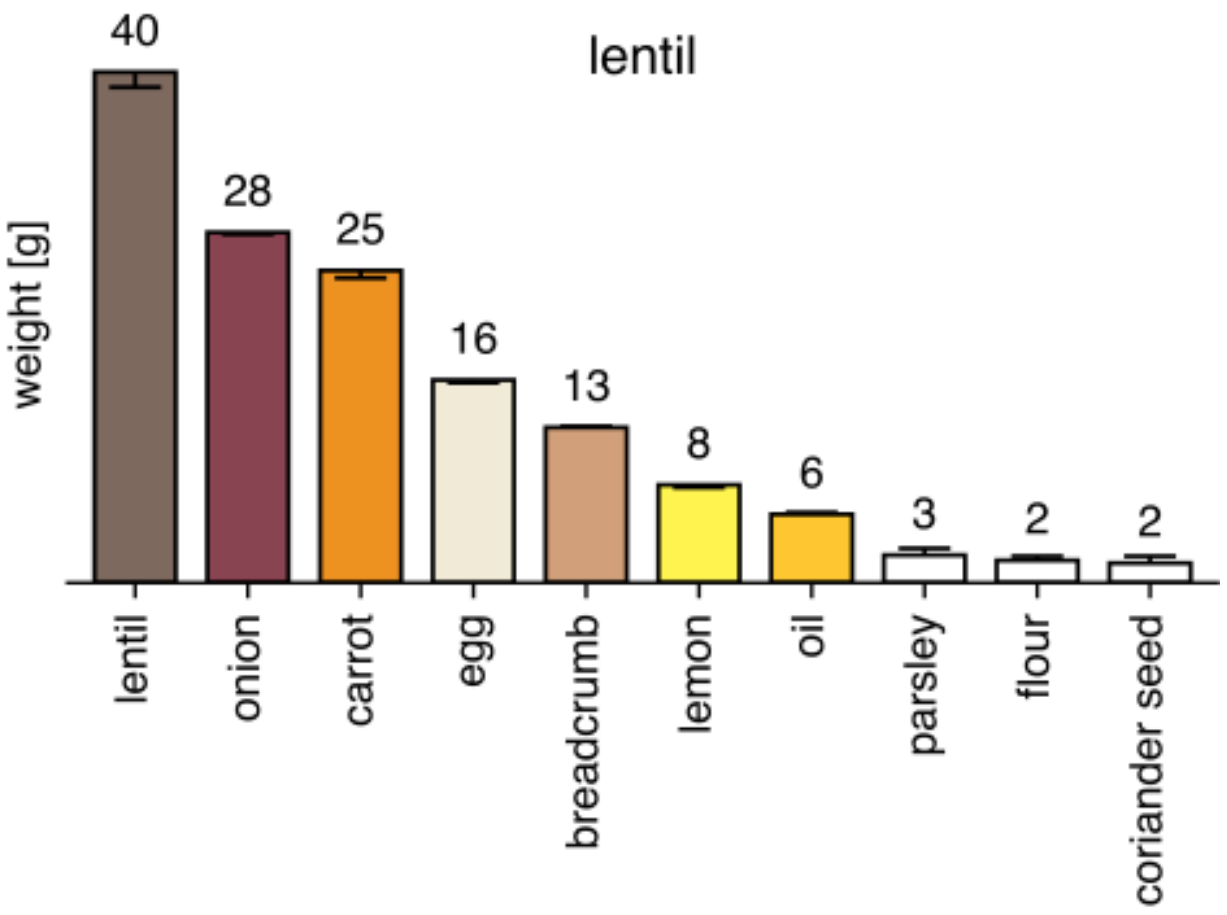


**Prep:**

*Place the rinsed/drained lentils onto a lined sheet tray and place in a 300°f oven for 3-5 minutes to dry off excess moisture*, careful not to overdry.

Combine all ingredients except ¼ teaspoon of the olive oil, ¼ tsp of the lemon juice, and the sliced onion and whole parsley leaves.

Mash the mixture together with a spatula or your hands until cohesive and form 30 equally sized balls (one per patty). Let rest for 10 minutes to hydrate.

Form into 30 x 113 g tightly packed patties and place in the fridge to firm up for 30 minutes.

Combine the remaining parsley leaves and sliced onion with the reserved olive oil and lemon juice. Weigh the total amount and divide by 30 to determine the total g amount per portion. Measure out into 30 individual portions and put in separate, labeled bags (one per burger, 30 total).

**Day of:**

Heat a lightly greased nonstick pan over medium heat (about 350°F) and cook on each side for 3-4 minutes and internal temp reaches 165°f

Serve the lentil patty on a toasted sesame seed bun with the parsley-onion garnish.

# Mushroom Burger

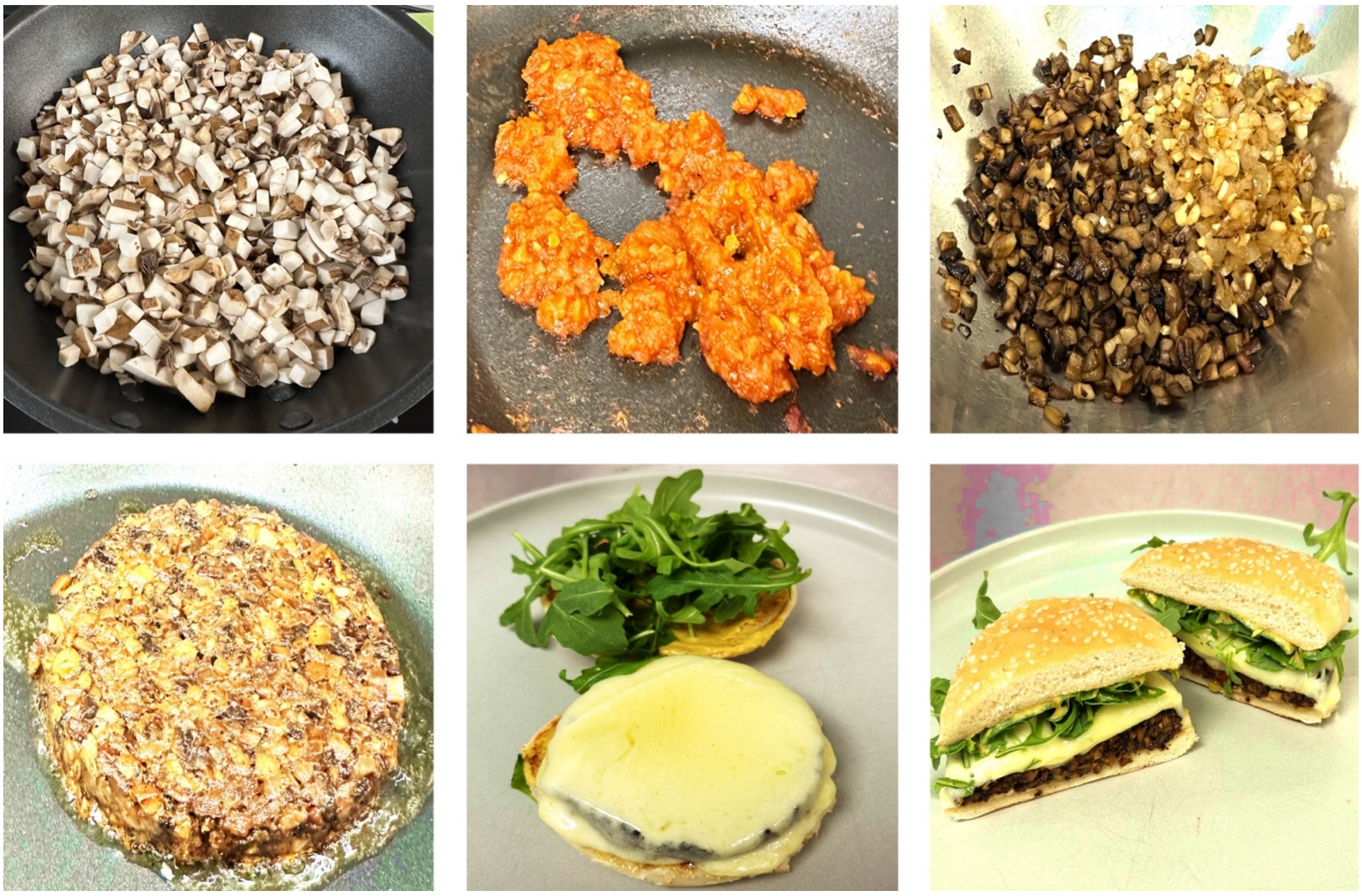

**Figure S4. Preparation of the mushroom burger.** Representative stages show ingredient preparation and mixing (top), the cooked bean patty, burger assembly, and the final burger in cross-section (bottom).

**The Mushroom Burger**

**Per burger:**

134 g Portobello Mushroom (stemmed and gills removed)
95 g Tomato, sliced
53 g Provolone cheese (pre-sliced)
37 g Onion, sliced
26 g Balsamic Vinegar
17 g Dijon Mustard
17 g Olive Oil
9 g Lettuce, gem or romaine
4 g Garlic (microplaned or minced finely)
3 g Pepper, ground black
1 g salt

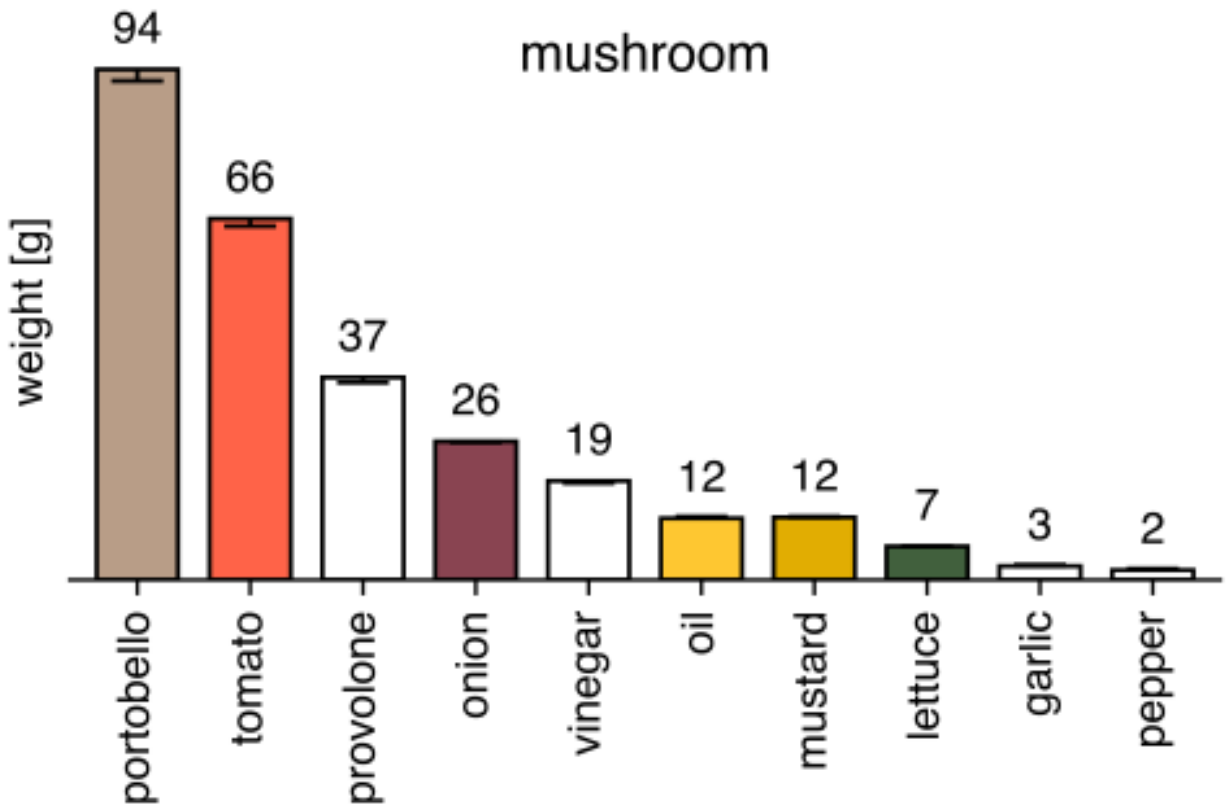


**Prep:**

Stir together the balsamic vinegar, mustard, garlic, salt, and pepper, then whisk in the olive oil to form the marinade. Put into 30 separate labeled ziplock bags, one for each burger.

Prepare the mushrooms by removing the stems and gills; peeling the cap is optional but helps the marinade penetrate.

Place the 134 g caps in the marinade baggies (134 g per burger into each baggie), coating them well, and refrigerate.

Pre-measure provolone cheese slices (53g each) and put into a labeled plastic bag, one for each burger (30 total bags).

Pre-measure onion (37g per burger), tomato (95g slice per burger), lettuce (9g per burger) and put combined into a labeled plastic bag, one for each burger (30 total bags).

**Day of:**

Sear the mushrooms in a non-stick pan over medium-high heat for about 3 minutes per side.

Transfer to a sheet tray, stack one cap on top of another, add cheese, and finish in a 350°F oven for a few minutes until melted.

Build the burger.

Alternatively, for larger batches, arrange the caps on sheet trays and bake in a conventional oven at 375°F for 20–25 minutes, or in a convection oven at 350°F for 10–15 minutes, until tender, then stack 2 or whatever weighs close to 113g and melt the cheese on top.

# Oat Burger (Savory)

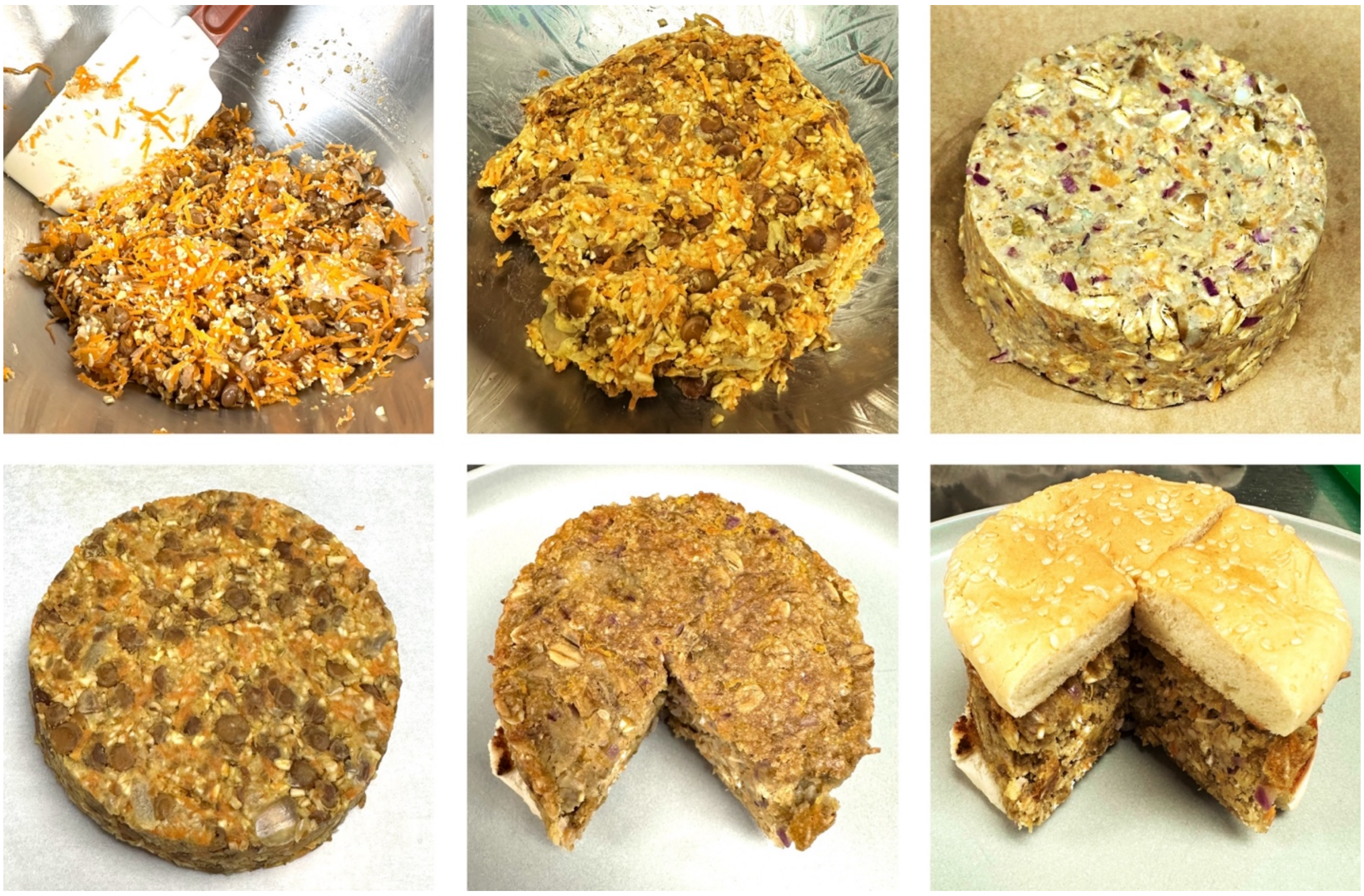

**Figure S5. Preparation of the savory oat burger.** Representative stages show ingredient preparation and mixing (top), the cooked bean patty, burger assembly, and the final burger in cross-section (bottom).

**5. The Savory Oat Burger**

**Per burger:**

29 g Oat, quick steel cut
40 g Lentil
18 g Carrot
15 g Onion
6 g Soy Sauce
2 g Olive Oil
3 g Breadcrumbs
1 g salt

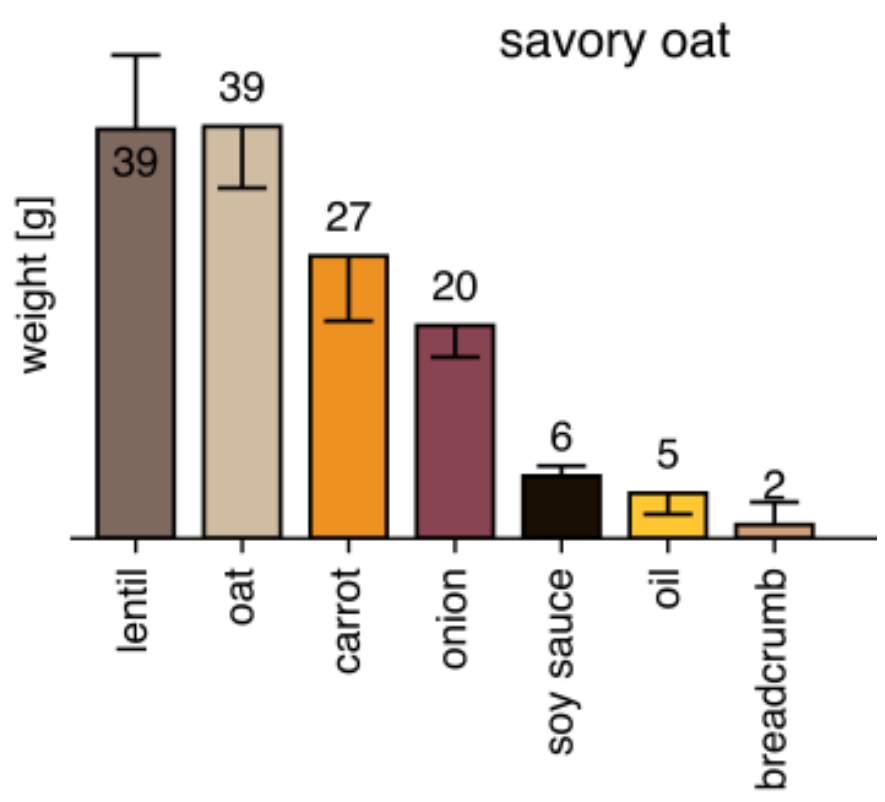


**Prep:**

Spread rinsed & drained lentils on a tray and warm them at 350°F for 3-5 minutes until the surface is dry but not toasted.

Grate the carrot and squeeze out excess moisture. Sweat the onion in a pan for 2–3 minutes until translucent, then cool.

Combine all ingredients and mash together until cohesive.

Let the mixture sit for 10 minutes so the oats and panko absorb moisture.

Form into 30 x 113 g patties and refrigerate.

**Day of:**

Sear over medium-high heat with a thin film of oil for 3–4 minutes per side until well browned and crisp.

# Oat Burger (Sweet)

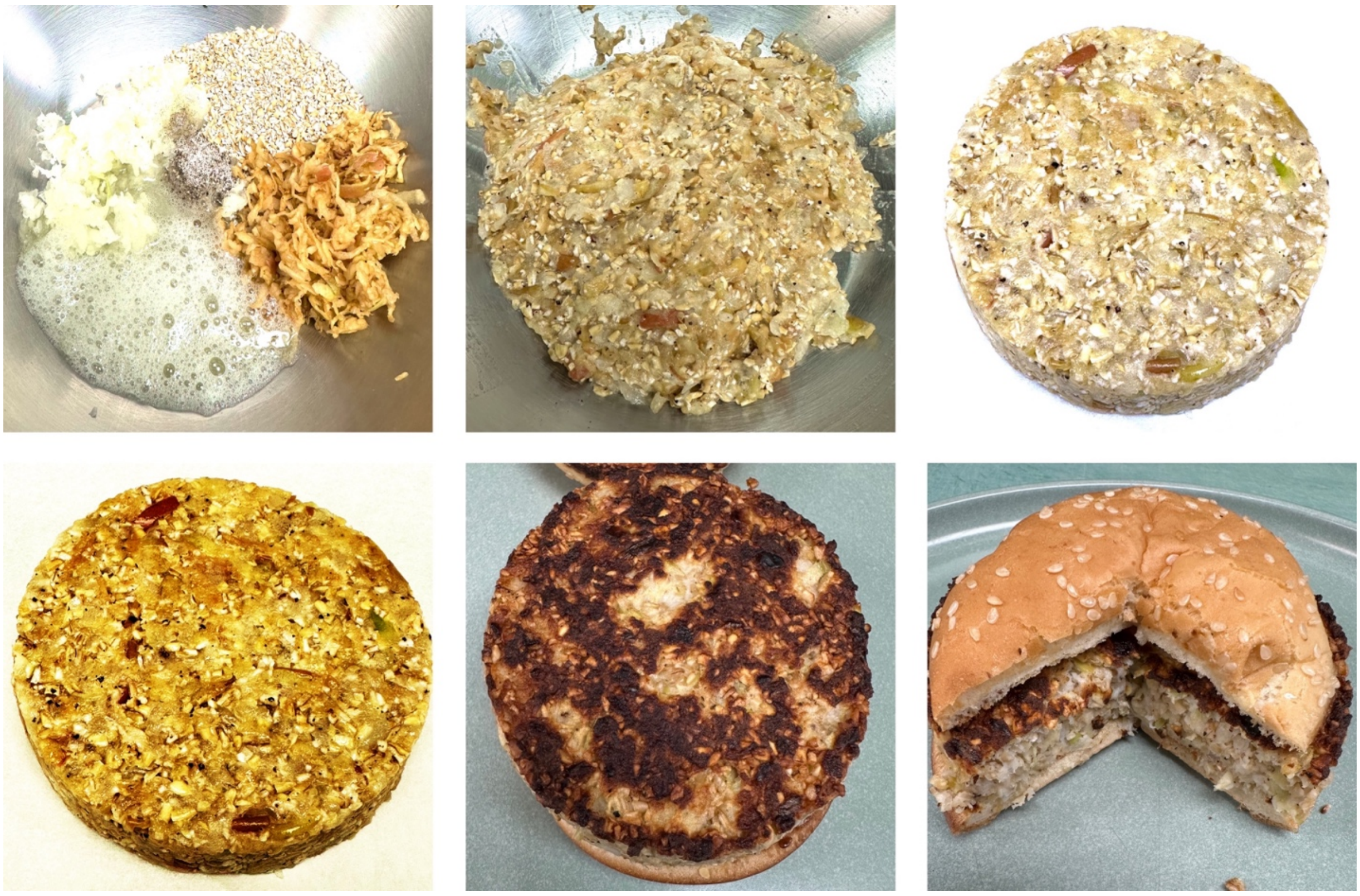

**Figure S6. Preparation of the sweet oat burger.** Representative stages show ingredient preparation and mixing (top), the cooked bean patty, burger assembly, and the final burger in cross-section (bottom).

**The Sweet Oat Burger**

**Per burger:**

29.4 g Quick Steel Cut Oats
35.4 g Egg whites
21.2 g Apple (any apple, not green)
26 g Yellow Onion
1 g Coarse cracked, black pepper
1 g salt

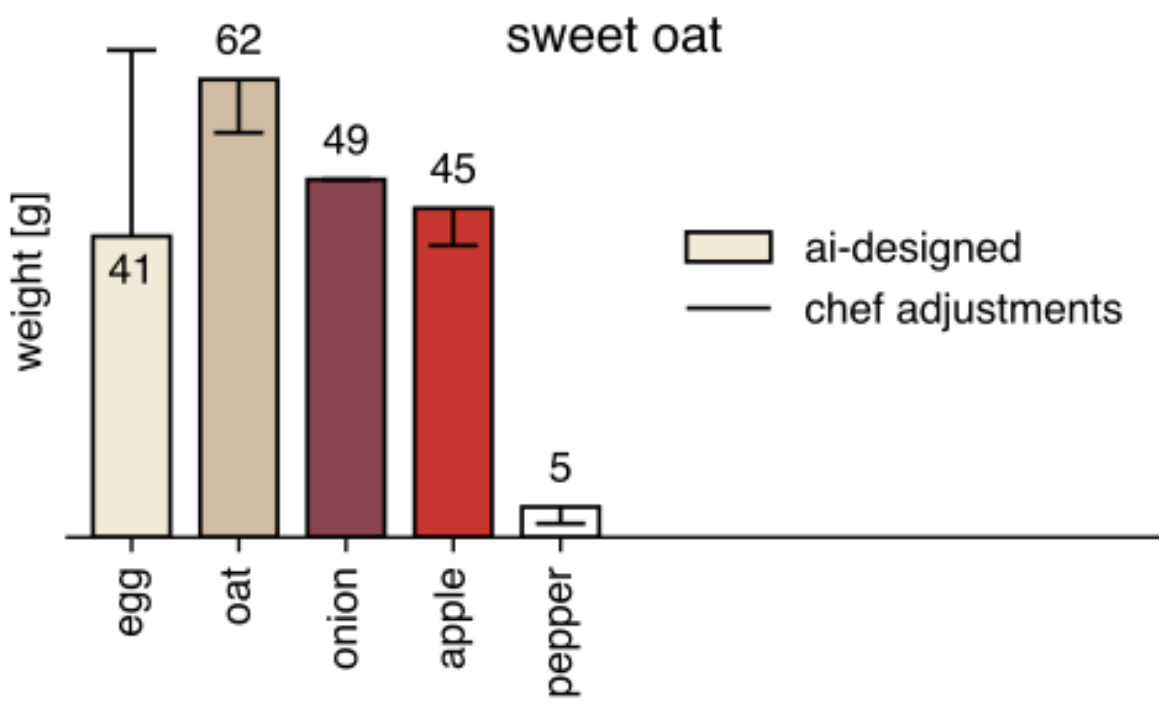


**Prep:**

Grate the apple and onion, wrap each in a towel or cheesecloth and squeeze hard to remove moisture. Weigh after squeezing.

Mix everything together well.

Let the mixture rest for 10 minutes so the oats can absorb the moisture.

Next, form into 30 x 113 g patties and chill.

**Day of:**

Sear over a medium high heat for 3 - 4 minutes per side. 165°f

# Restaurant Survey

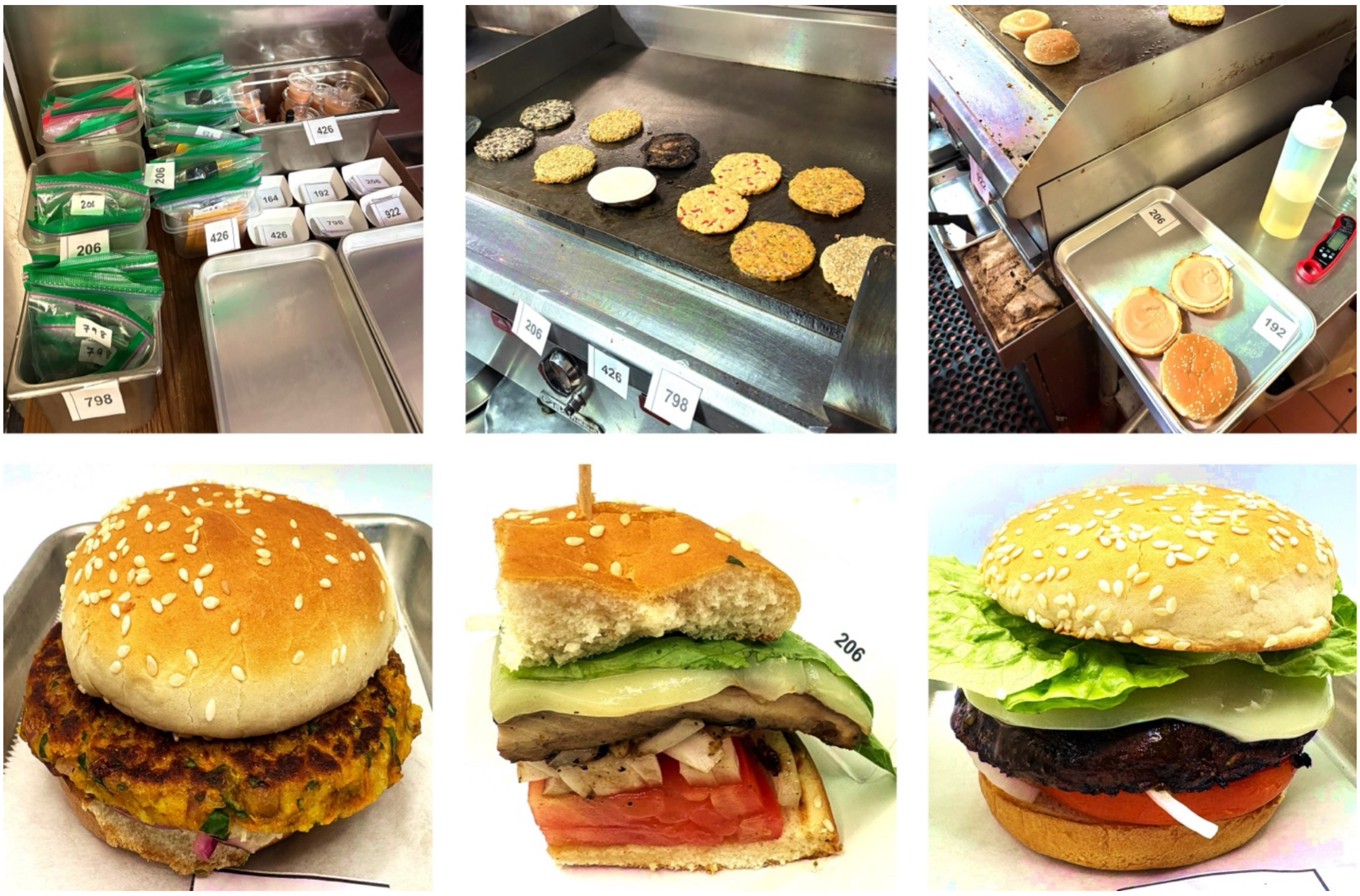


**Figure S7. Scenes from the restaurant survey.** Preparation of the burger patties (top) and selected burgers in cross-section (bottom).

We conduct the blinded sensory survey with 101 participants at a restaurant in San Francisco. Restaurant chefs prepare all six burgers according to standardized protocols and serve them fresh. Each participant evaluates the burgers in a randomized order without knowledge of their identities or ingredients. Figure S7 documents the restaurant kitchen, preparation of the burger patties, and representative cross-sections of the prepared burgers.


\end{document}